\documentclass{elsart}
\usepackage{epsf}

\begin{document}
\begin{frontmatter}

\title{Alternation of different fluctuation regimes in the stock 
market dynamics}

\author{J. Kwapie\'n$^{1,2}$, S. Dro\.zd\.z$^{1-3}$ and J. Speth$^1$}

\address {$^1$Institut f\"ur Kernphysik, Forschungszentrum J\"ulich,
D-52425 J\"ulich, Germany \\
$^2$Institute of Nuclear Physics, PL--31-342 Krak\'ow, Poland \\
$^3$Institute of Physics, University of Rzesz\'ow, PL--35-310 Rzesz\'ow, 
Poland}

\begin{abstract}

Based on the tick-by-tick stock prices from the German and American stock
markets, we study the statistical properties of the distribution of the
individual stocks and the index returns in highly collective and noisy
intervals of trading, separately. We show that periods characterized by
the strong inter-stock couplings can be associated with the distributions
of index fluctuations which reveal more pronounced tails than in the case
of weaker couplings in the market. During periods of strong correlations 
in the German market these distributions can even reveal an apparent 
L\'evy-stable component.

\end{abstract}

\end{frontmatter}

\section{Introduction}

A series of papers devoted to the analysis of financial data fluctuations
disclosed that the corresponding distributions can be characterized by the
Paretian scaling~\cite{mandel63,mant95,lux96,plerou99b,gopi99}. These
studies, based on the large data sets of historical stock prices and on
the index values, showed that both the distributions of stock price
fluctuations and the distriutions of index returns reveal scaling over a
broad range of time scales from minutes to days (although a more recent
investigation found that scaling is restricted to rather short time
scales~\cite{drozdz02b}). A remarkable related issue is that the stocks
and indices exhibit similar value of the scaling exponent $\alpha \simeq
3.0$~\cite{plerou99b,gopi99}. In accordance with the Central Limit
Theorem, the distribution of a random variable being a sum of a number of
{\it iid} random variables with a finite second moment, has to converge to
a normal distribution. From this point of view, the similarity of the
distributions for the stocks and the corresponding indices requires that
the financial data violate the assumptions of the theorem. And as these
data have indeed finite variance, a plausible cause for the problems with 
the convergence can be related to the correlations among the data. This 
claim seems to be supported by findings that an artificial S\&P index 
constructed from randomly reshuffled stock returns, presents a much better 
convergence to a Gaussian than the original index~\cite{plerou99b}.

An appropriate measure of correlations among elements of a system is the
spectrum of the correlation matrix eigenvalues, which can be easily
compared with the universal properties of random matrices~\cite{mehta91}.
A few recent works have shown that the financial market can be described
by at least one repelled eigenvalue with a magnitude exceeding the likely
range of values allowed for a random matrix. This one or more deviating
eigenvalues indicate that there are relations between various components
of the market~\cite{laloux99,plerou99a,drozdz00,drozdz01a}.

The main purpose of the present work is a quantitative description of the
possible relation between the stock price movements and the properties of
the distribution of the corresponding index fluctuations. We showed in a
previous analyses which were focused on daily patterns of the German DAX
index fluctuations that certain characteristic time intervals of a trading
day with high index volatility are associated with fluctuation
distributions with properties different from more silent intervals of
trading~\cite{drozdz01b,kwapien02}. Since high volatility is connected
with stronger correlations between the stocks (\cite{cizeau01,mounf01}) we
expect that strong and weak inter-stock correlations are reflected in
different properties of the index fluctuations. By choosing a few distinct
time scales (1-30 minutes) we are able to test the stability of the
results.

\section{Methodology}

Our analysis is based on the high frequency tick-by-tick data covering
the two years 1998-99 period and comprising the recordings of 30
companies included in the Dow Jones Industrial Average and 30 companies
included in the German DAX30 index, together with the two 
indices~\cite{data}. Inevitably, such a long interval of time comprises
some changes of the index composition. We decided that only those stocks
which were a part of an index for the majority of time, can be taken
into consideration. Along this way, for the whole interval under study
we analyze the data for the individual companies CHV, GT, S and UK, 
although on Nov 1, 1999 they were replaced in DJIA by HD, INTC, MSFT and 
SBC. In a similar manner for the German market, we analyze ADS instead of 
BVM (delisted due to its fussion with BHW). For the German stock market we 
have roughly 30\% more data points, because of a longer trading day in 
Frankfurt (8:30 hours vs. 6:30 hours in New York). The data has been
preliminary processed to clear out recording errors.

Let us assume we have a set of $N$ assets and $x_{\beta}(t_i)$ 
($i=1,...,T$) is a price of the $\beta$-th asset at instant $t_i$. The 
corresponding time series of normalized logarhitmic price returns reads:
\begin{equation}
g_{\beta}(t_i)=\frac {G_{\beta}(t_i) - \langle G_{\beta}(t_i) 
\rangle_{t_i}} {\sigma(G_{\beta})}, \ \quad 
\sigma(G_{\beta}) = 
\sqrt{\langle G_{\beta}^2(t_i) \rangle_{t_i} - \langle 
G_{\beta}(t_i) \rangle_{t_i}^2}
\end{equation}
where
\begin{equation}
G_{\beta}(t_i) = \ln x_{\beta}(t_i+\Delta t) - \ln x_{\beta}(t_i).
\end{equation}
The time lag $\Delta t$ defines a time scale and $\langle \ldots 
\rangle_{t_i}$ stands for averaging over discrete time.
From all the time series $g_{\beta}(t_i)$ ($\beta=1,...,N; \ i=1,...,T$)
we construct an $N \times T$ data matrix ${\bf M}$ and then calculate a 
correlation matrix ${\bf C}$ defined by
\begin{equation}
{\bf C} = (1 / T) \ {\bf M} {\bf M}^{\rm T},
\end{equation}
which is is an $N \times N$ square matrix with correlation coefficients as 
its entries.

After the correlation matrix is calculated, we diagonalize it and obtain a
spectrum of its eigenvalues $\lambda_k$ ($k=1,...,N$). For a random matrix
(the so-called Wishart matrix) constructed from series of random numbers
taken from the normal distribution, in the limit of $N \rightarrow \infty$
there exist exactly $N$ non-zero eigenvalues, providing $Q:=T/N>1$ (and
$N-1$ ones for $Q=1$). In this case an analytic expression for the
distribution of the matrix eigenvalues exists~\cite{seng99}:
\begin{eqnarray} 
\label{eq:rho} 
\rho_C(\lambda) = {Q \over {2 \pi \sigma^2}} {\sqrt{ (\lambda_{\rm max} -
\lambda) (\lambda - \lambda_{\rm min})} \over {\lambda}}, \\ \centering
\lambda^{\rm max}_{\rm min} = \sigma^2 (1 + 1/Q \pm 2 \sqrt{1/Q}),
\nonumber \end{eqnarray} with $\lambda_{\rm min} \le \lambda \le
\lambda_{\rm max}$,
and where $\sigma^2$ is equal to the variance of the time series. In both
cases, any deviation from the universal Random Matrix Theory predictions
means that the correlation matrix comprises some genuine information
specific for the system under study.

In a system like the stock market, the correlation matrix usually reveals
at least one strongly repelled eigenvalue, describing the common behaviour
of a group of assets or even the common evolution of the whole
market~\cite{laloux99,plerou99a,drozdz00,gopi01,plerou02}. The magnitude
of such an eigenvalue is related to the range of correlation of different
asset prices. A more collective market behaviour is reflected in a larger
$\lambda_1$. In the real markets the range of the inter-stock couplings
turns out to be strongly time-dependent; this refers both to long time
scales (daily returns)~\cite{drozdz00,onnela03} and to extremely short
ones (of order of minutes and even seconds)~\cite{drozdz01b,kwapien02}. As
it has already been mentioned, the correlations grow in highly volatile
periods of time and fade during more silent intervals, when trading is
dominated by noise (see~\cite{drozdz00}, but also~\cite{onnela03}). The
sudden elevation of the largest eigenvalue in crash times and decrease in
onset periods of a rally can serve as examples of such behaviour.

A possible influence of the asset coupling strength on the distribution
of index returns may be observed by comparing returns corresponding to
intervals with strong correlations (large values of $\lambda_1$) and
those from intervals with weak correlations (small $\lambda_1$). To
accomplish this, we divide the whole two-years-long period under study
into equal disjoint time windows $w_j$, $j=1,...,n_w$ each of length
$T_w$. In each window we calculate the correlation matrix ${\bf C}$ and
its largest eigenvalue $\lambda_1(w_j)$. Next we determine the eigenvalue
distribution $P(\lambda_1(w_j))$ and select such windows $w_k$ that
$\lambda_1(w_k)$ falls within a specific range of values of the
$P(\lambda_1(w_j))$ distribuant. Finally, we compute the distributions of
index and stock returns belonging to these selected windows. Since we
need a good time resolution for fixed $N=30$, we choose $T_w=30$ (i.e.
$Q=T_w/N=1$), regardless of the time scale $\Delta t$. From the RMT
perspective this fixes the upper edge of the random eigenvalues bulk at
exactly $\lambda_{\rm max}=4.0$. However, since we use very short windows
($N\ll\infty$), the average value of $\lambda_1$ may be smaller than
$\lambda_{\rm max}$ in the random case and $\lambda_1(w_j)$ may fluctuate
around the average for different $j$. (Fortunately, these fluctuations
described by the so-called Tracy-Widom distribution~\cite{tracy97,john00}
are relatively small already for $N=30$ and thus do not influence our
findings for $\lambda_1 \gg \lambda_{\rm max}$.) For our analysis, four
distinct time scales were chosen: $\Delta t=1, \Delta t=5, \Delta t=10$
and $\Delta t=30$ minutes; we could not include any higher scales due to
extremely poor statistics of returns in that case.

\section{Results}

\begin{figure}
\epsfxsize 11cm
\hspace{1.0cm}
\epsffile{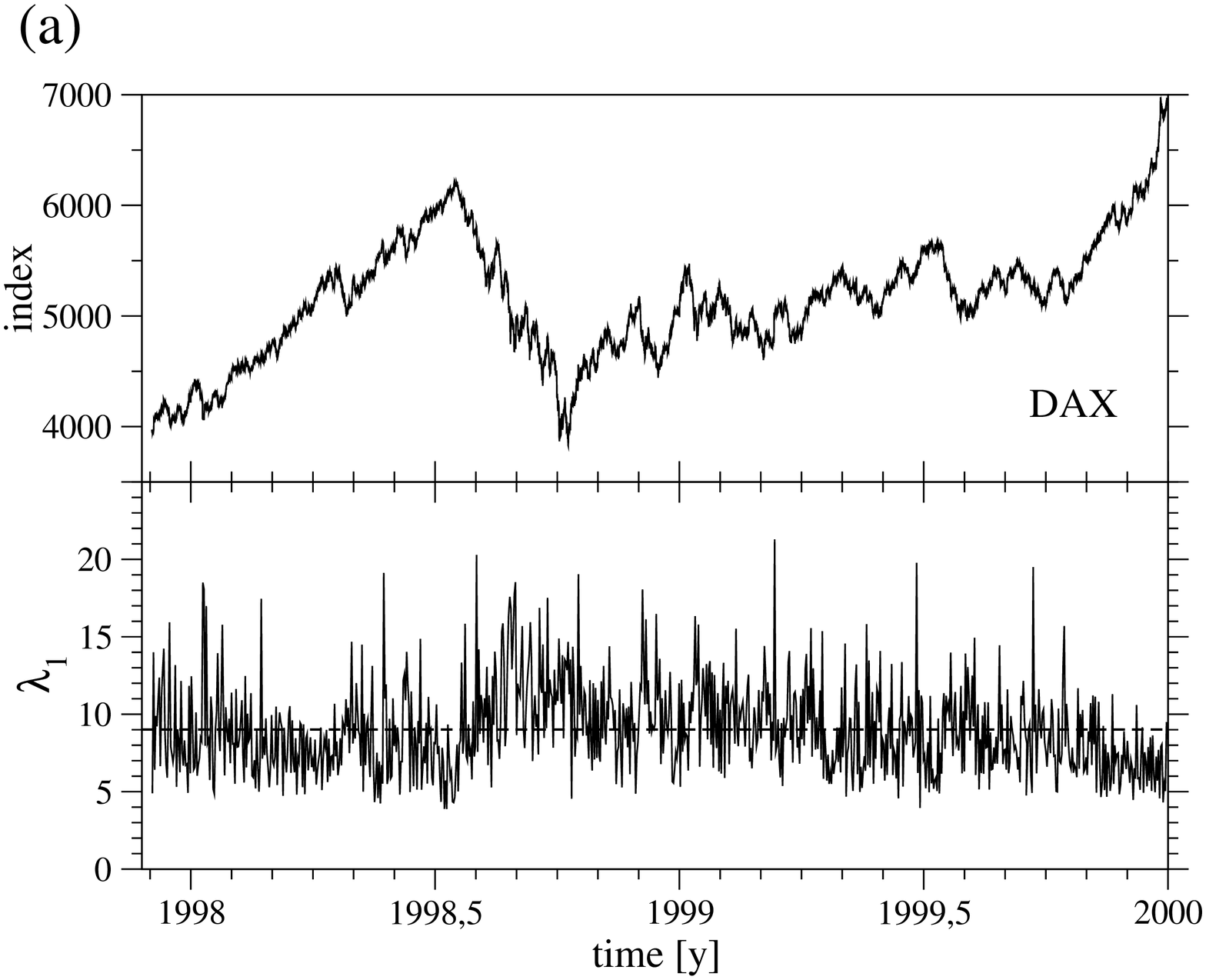}

\vspace{0.2cm}
\hspace{1.0cm} 
\epsfxsize 11cm
\epsffile{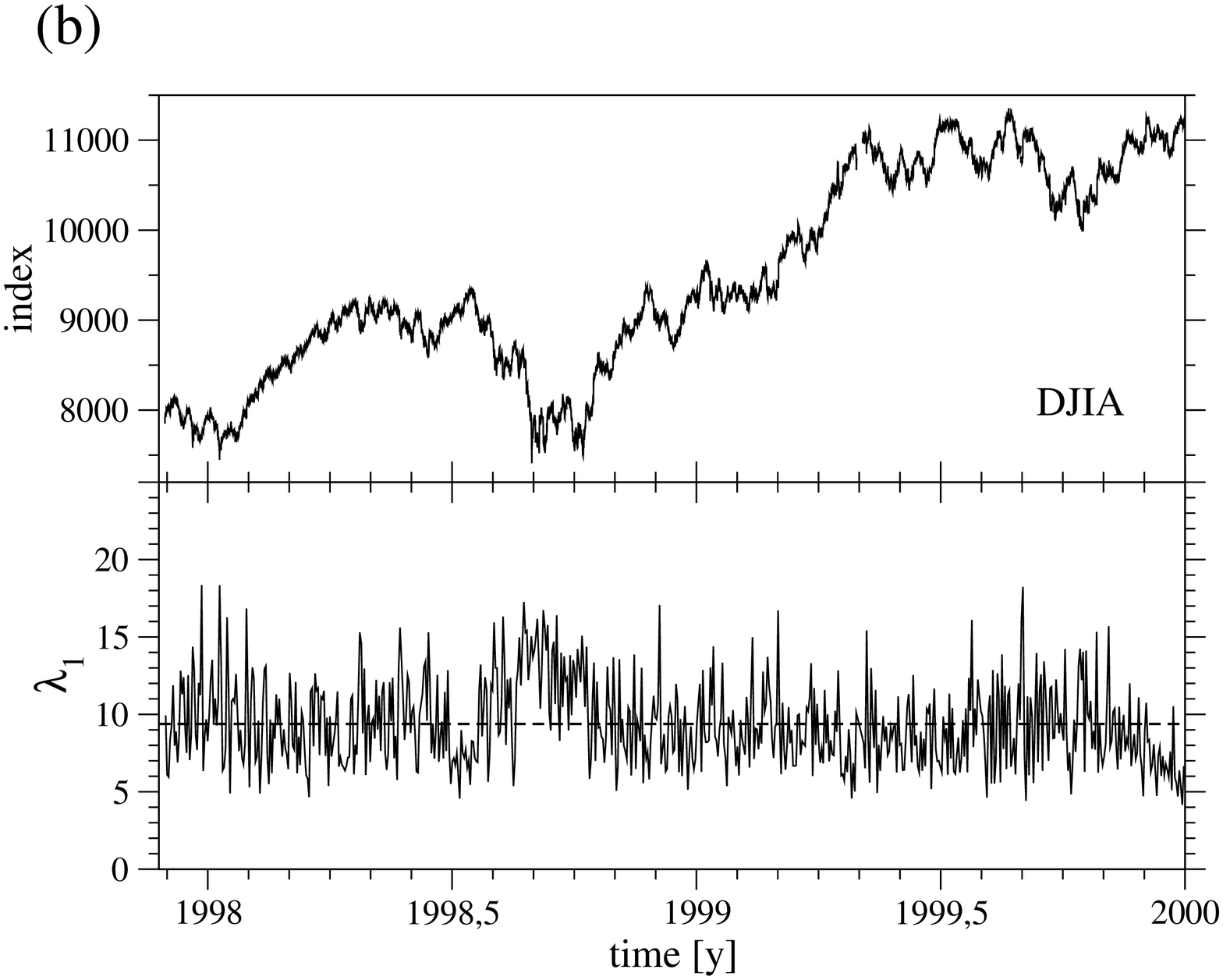}
\caption{Time course of the DAX index (a) and of the DJI Average   
(b) sampled with 10 min frequency (upper panels) together with the     
corresponding $\lambda_1(w_j)$-values calculated in time windows of 
$T_w=30$ data points (lower panels). The arithmetic average of the 
eigenvalue over all the windows is also indicated (dashed line).}
\label{fig:fig1}
\end{figure}

The upper panels of Figure 1 present the time course of the DAX (a) and of
the DJIA (b) indices (here sampled every 10 minutes) in the studied time
interval, i.e. between Dec 1, 1997 and Dec 31, 1999. Over this period,
both indices experienced a significant increase: $\simeq$ 70\% in the case
of DAX and $\simeq$ 40\% in the case of DJIA, although large drowdowns
were also observed (e.g. in August 1998). In principle, however, that was
exactly this period in which the most powerful bull market was observed
both at NYSE and at Deutsche B\"orse. For a comparison, the lower panels
display the largest eigenvalue for each time window $\lambda_1(w_j)$ as a
function of time; each point corresponds to a single time window $w_j$
which is 300 min (30 data points) long in the present case. The varying
degree of collectivity is clearly visible here. This resembles the
evolution of $\lambda_1$ calculated from daily data in
ref.~\cite{drozdz00}. For the sake of clarity of the Figure, we chose
$\Delta t=10$ min, since smaller time scales are associated with more
noisy dynamics. Comparing the lower panels of (a) and (b), we see that
both the markets have similar average value of $\lambda_1$ (denoted by
horizontal lines) equal to 9.01 (DAX) and 9.38 (DJIA), respectively. This,
however, cannot be treated as a rule, because for shorter $\Delta t$ DAX
is on average significantly more collective than DJIA. Even for this time
scale of 10 min the German market develops $\lambda_1$ which reaches
extremely high values significantly more often than it happens in the Dow
Jones market (compare ref.~\cite{drozdz00}). The difference between the
values of $\lambda_1$ in the strongly collective and the least collective
periods of time is striking. The smallest $\lambda_1(w_j)$'s fall inside
the noisy part of the eigenvalue spectrum ($\lambda_1 \simeq \lambda_{\rm
max}$) suggesting that no meaningful correlations are present at that
time, while, on the other hand, the largest $\lambda_1(w_j)$'s, almost
saturating the available range of values ($\lambda_1 \le {\rm Tr} {\bf
C}=N$), describe nearly ``rigid'' market.

In order to compare the statistical properties of the stock and of the 
index returns across windows with different degree of inter-stock 
couplings, we introduce a pair of parameters $\zeta^W$ and $\zeta^S$ 
defined by the following relation
\begin{equation}
\zeta^{W,S} := \frac {\#\{w_j: \lambda_1(w_j) < \Lambda^{W,S}\}} 
{n_w}, \ \ (\zeta^W \le \zeta^S),
\label{eq:zeta}
\end{equation}
where $\Lambda^S$ denotes the lower threshold for $\lambda_1$, defining 
the strongly correlated market and $\Lambda^W$ denotes the upper 
threshold when the market is considerably weakly correlated. Specific 
values of $\Lambda^W,\Lambda^S$ depend on a particular choice of 
$\zeta^W,\zeta^S$. By this definition, the case of $\zeta^W = \zeta^S = 
0.5$ corresponds to the median of the distribution, while the one of 
$\zeta^W = 0.2, \zeta^S = 0.8$ to the 20th and 80th percentile of the 
distribution, respectively. From now on, we will recognize two cases: 
periods of collective trading (strong stock cross-correlations, $S$) if 
$\lambda_1(w_j) > \Lambda^S$ and periods of uncorrelated trading (weak 
cross-correlations, $W$) if $\lambda_1(w_j) < \Lambda^W$.

We anticipate that each of these two cases is represented by a
distribution of index returns with somewhat different properties. An
index can be either a simple sum of the related stock prices (e.g. Dow
Jones) or a sum of prices weighted by capitalization of the
corresponding companies (e.g. DAX and S\&P indices family). This
suggests that the index fluctuations may be described by a distribution
being closer to a Gaussian in periods of uncorrelated trading, when a
lack of profound correlations brings situation close to the assumptions
of the Central Limit Theorem, while being significantly different from
normal during strongly collective stock behaviour, when these
assumptions are firmly violated. In contrast, the properties of the
corresponding distributions for the individual stock returns may not
be so sensitive to the inter-stock correlations as the index returns
are.

\begin{figure}
\epsfxsize 11cm
\hspace{1.0cm}
\epsffile{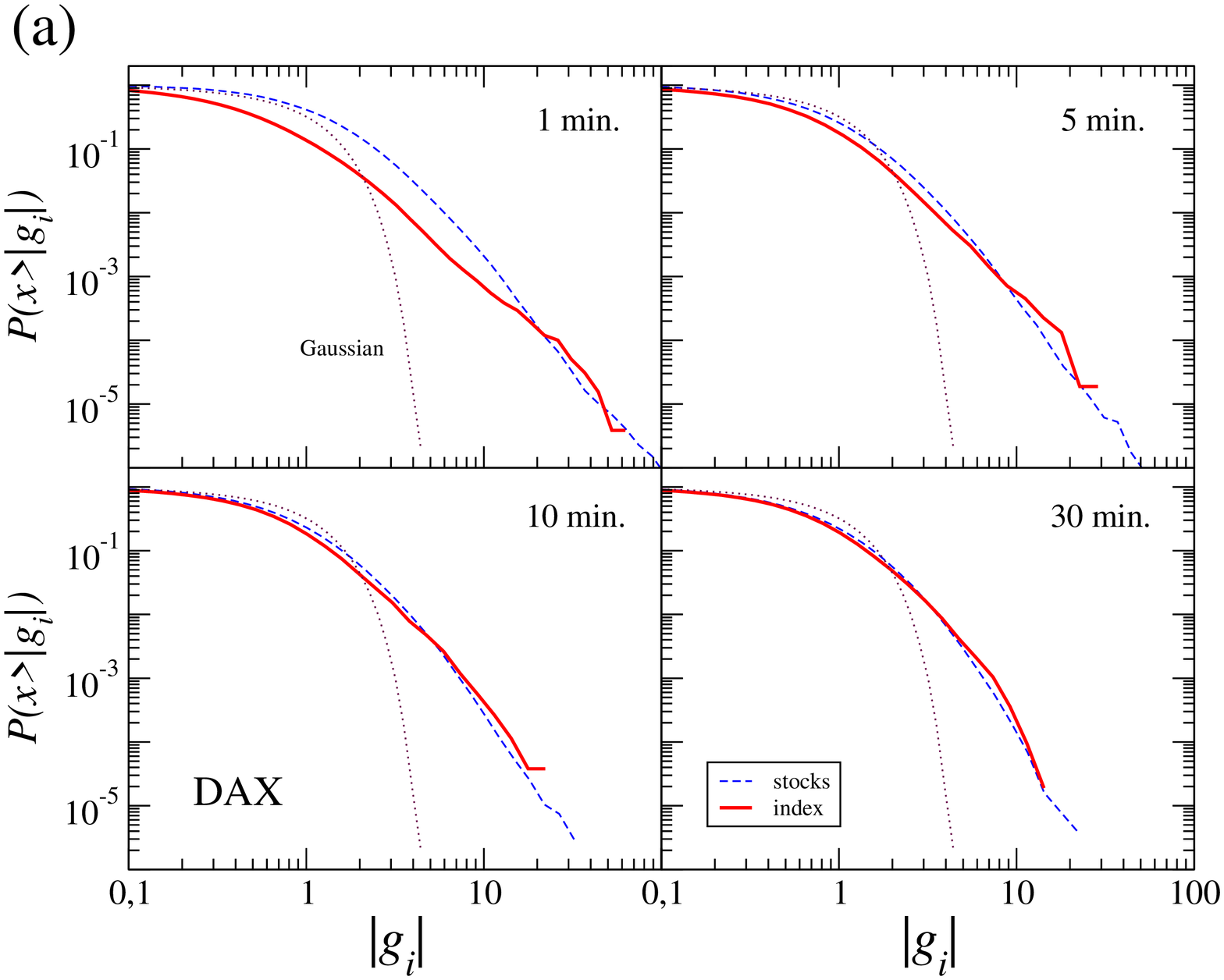}

\vspace{0.8cm}
\epsfxsize 11cm
\hspace{1.0cm} 
\epsffile{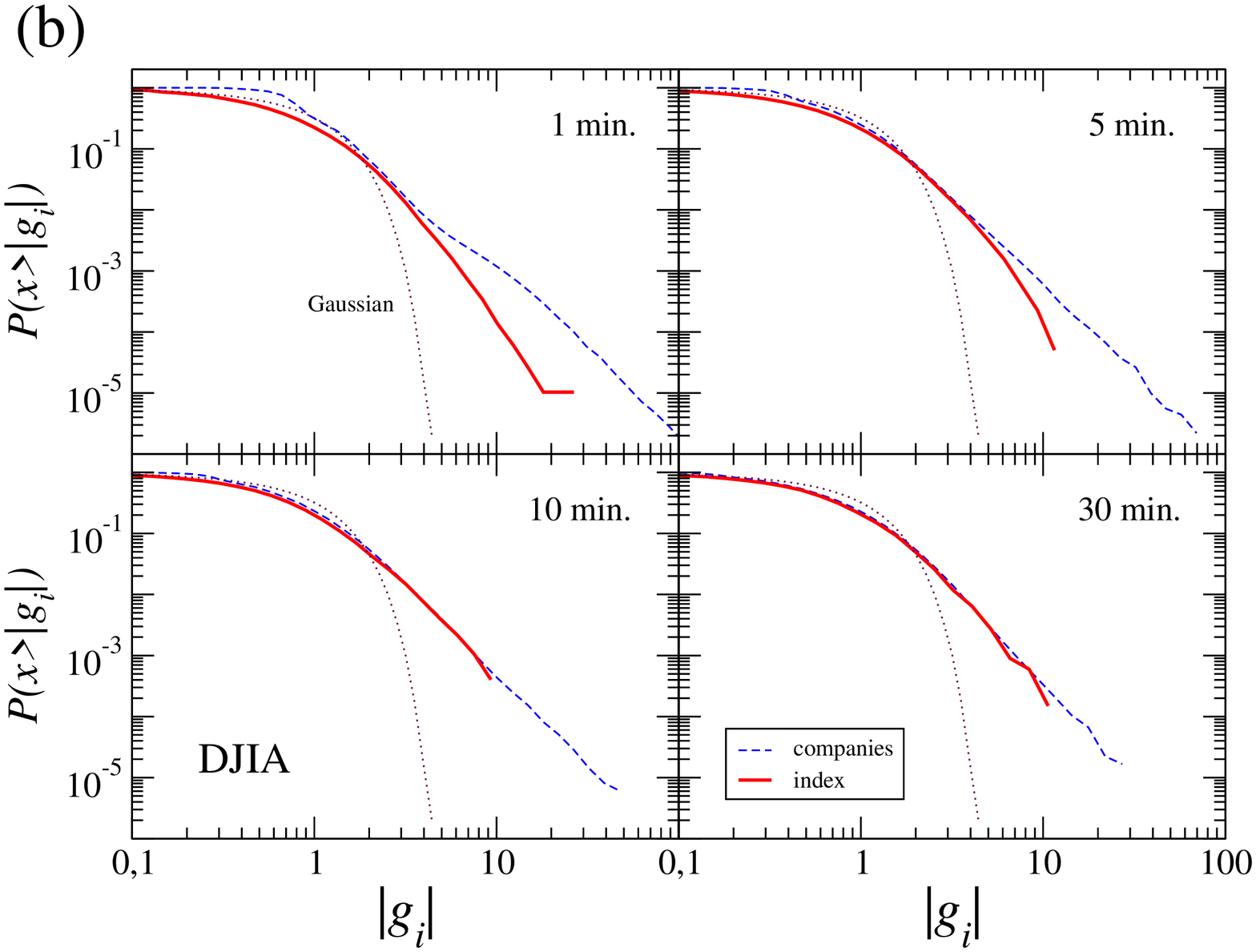}
\caption{Cumulative distributions of normalized stock returns (dashed) and
of normalized index returns (solid) for four different time scales (1-30
min) and the two markets studied: DAX (a) and DJIA (b). The cumulative
normal distribution is indicated by a dotted line in each panel.}
\label{fig:fig2}
\end{figure}

Before we divide the signals into the correlated and noisy parts, we show
in Figure 2 the cumulative distributions of the normalized stock returns
and of the index returns for the DAX market (Fig.~2(a)) and the DJI market
(Fig.~2(b)). This Figure allows one to compare the distribution of the
stock returns calculated for all 30 companies constituting each of the
markets (dashed line) with the distribution of the index returns (solid
line), for four different $\Delta t$. As it can be seen in Figure, the
distributions of the stock returns scale in their tails for both the
markets and for all the time scales. For the two shortest time scales the
tails of the distributions of the index returns for DAX and for DJIA have
different properties, however. On one hand, in DAX, the slope of the
distribution of the index returns is smaller than the slope of the
corresponding distribution of the stock returns. On the other hand, the
reverse effect is observed for DJI stocks and DJIA. Interestingly, by
increasing $\Delta t$ from 1 to 30 min the difference between the index
returns distribution and the associated stock returns distribution
disappears for both DAX and DJIA. Nevertheless, in each case in Fig.~2 all 
the distributions are far from Gaussian.

Later on, we shall discuss the origin of the difference in the slope of
the tails between the DAX and the DJIA returns distributions on short time
scales in more detail. Here we only indicate that among the sources of
discrepancy between the stock and the index distributions is the existence
of periods when the price of a stock does not change. This situation is
specific only to very short time scales and results in a number of zero
returns in the data. Such returns, of course, influence the average
volatility of price changes and thus also affect the normalization, giving
broader distributions of the returns while not affecting the tail's slope.
For the indices, however, this effect can be neglected even at 1-minute
time scale.

The outlying points in the distributions, being especially evident in DJIA
for 1 min returns (upper left panel of Fig.~2(b)) but also to a lesser
extent in DAX, can be almost exclusively related to large jumps of the
price of a single asset which enters the index with a significant weight.
Such jumps occur occasionally after some important information reaches the
market (financial reports, company fussions, takeovers etc.); the two
extreme examples are the 15\% increase and the over 20\% decrease of the
IBM share price on Apr 22, 1999 and Oct 21, 1999 openings, respectively.
Due to the fact that these single-stock jumps alone caused about 1\%
change of the DJIA value, the event which is rather unusual in the
time-period analyzed, we decided to remove these IBM stock returns and the
two corresponding DJIA returns from our time series for all the time
scales. This does not influence the essential results of our analysis in
any case.

\begin{figure}
\epsfxsize 11cm
\hspace{1.0cm}
\epsffile{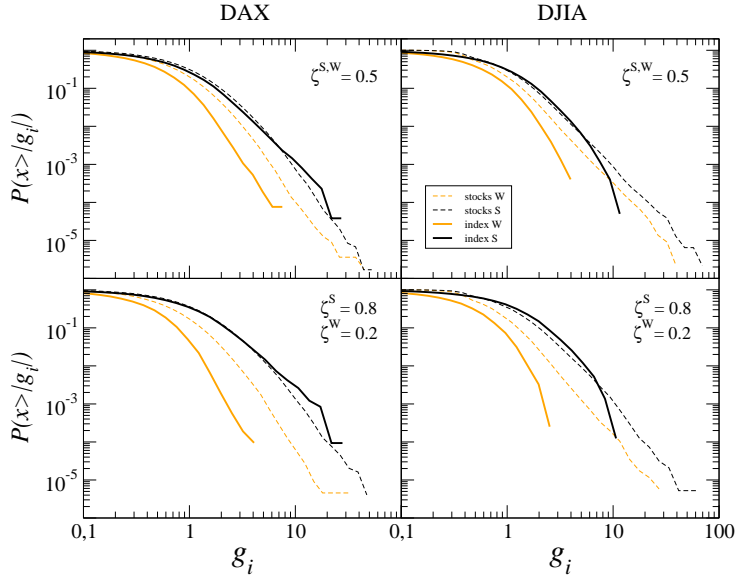}
\caption{Cumulative distributions of 5 min stock and index returns
for DAX (left column) and DJIA (right column) without the repeated
normalization. The distributions of returns corresponding to time 
windows of weak (W) and strong (S) correlations are presented together 
in each panel. The upper panels show distributions calculated for the 
returns from 50\% of the windows with highest value of $\lambda_1$ 
($\zeta^S=0.5$) and from the remaining 50\% of the windows with small 
$\lambda_1$ ($\zeta^W=0.5$). The lower panels display the same for 20\% 
of the windows with highest ($\zeta^S=0.8$) and smallest ($\zeta^W=0.2$) 
correlations. Gray dashed lines exhibit stock returns distributions for 
the W case and black dashed lines for the S case, while the corresponding 
index returns distributions are denoted by solid lines: gray (W) and 
black (S). Note that the two returns corresponding to IBM large jumps on 
Apr 22, 1999 and Oct 21, 1999 have been removed before calculation of the 
distributions both for the DJI stocks and for DJIA; the same refers to 
all subsequent figures.}
\label{fig:fig3}
\end{figure}

After calculating $\lambda_1(w_j)$ for all $j$'s, we define and select the
windows with strong and weak correlations between the companies. We
introduce three distinct kinds of window selection characterized by the
following values of the parameters: $\zeta^W=0.5$ and $\zeta^S=0.5$ (one
half of the windows are considered as covering weakly collective trading,
and the other half - strongly collective trading), $\zeta^W=0.2$ and
$\zeta^S=0.8$ (20\% of windows with smallest $\lambda_1$, 20\% of windows
with highest $\lambda_1$), $\zeta^W=0.05$ and $\zeta^S=0.95$ (5\%
smallest, 5\% highest).

Figure 3 displays the cumulative distributions of the stock returns
(dashed) and the index returns (solid) according to the local inter-stock
correlation strength ($W$ - gray, $S$ - black) for the DAX (left panels)
and for the DJI (right panels) markets. We present only one time scale
$\Delta t=5$ min, but these results are qualitatively stable across all
analyzed scales. For the companies, both the $W$ and $S$ distributions
(which we shall denote as $DAX_C^W$, $DAX_C^S$, $DJIA_C^W$ and $DJIA_C^S$)
have similar slopes with $DAX_C^W$ and $DJIA_C^W$ slightly shifted to the
left compared to $DAX_C^S$ and $DJIA_C^S$. This shift is caused by
different variance of the distribution in the $W$ and $S$ cases and it
illustrates the already known fact, that large price fluctuations
(volatility) are more likely to happen when the market is more collective
(and {\it vice versa})~\cite{cizeau01,mounf01}. In fact, the difference
between the distributions enlarges with increasing
$\Delta\zeta:=\zeta^S-\zeta^W$.

Another interesting conclusion can be drawn from the distributions of
index returns (these we denote as $DAX_I^W$, $DAX_I^S$, $DJIA_I^W$ and
$DJIA_I^S$). Here the differences between the $W$ and $S$ distributions,
for both of the indices, are amplified compared to the stock returns,
which is evident both for $\zeta^{W,S}=0.5$ and $\zeta^{W}=0.2$,
$\zeta^{S}=0.8$, with the separation much larger in the latter case. As
such a large difference cannot be explained merely on the basis of the
divergence between the associated distributions for the companies, the
range of correlations may be the crucial factor here.

Although some conclusion on the shape of the distributions in their tails 
can be drawn directly from the lower panels of Fig.~3, a real comparison 
may be performed only if all the sets of the returns are normalized once 
again to have a unit variance (so far the variance is different within 
each of the interrelated $W$ and $S$ sets). Figure 4 is the central point 
of the present paper. It exhibits the cumulative distributions of the 
stock and index returns after the repeated normalization. Parts (a)-(d) 
show the results for $\Delta t=1, 5, 10$ and $30$ min, respectively.

\begin{figure}
\epsfxsize 11cm
\hspace{1.0cm}
\epsffile{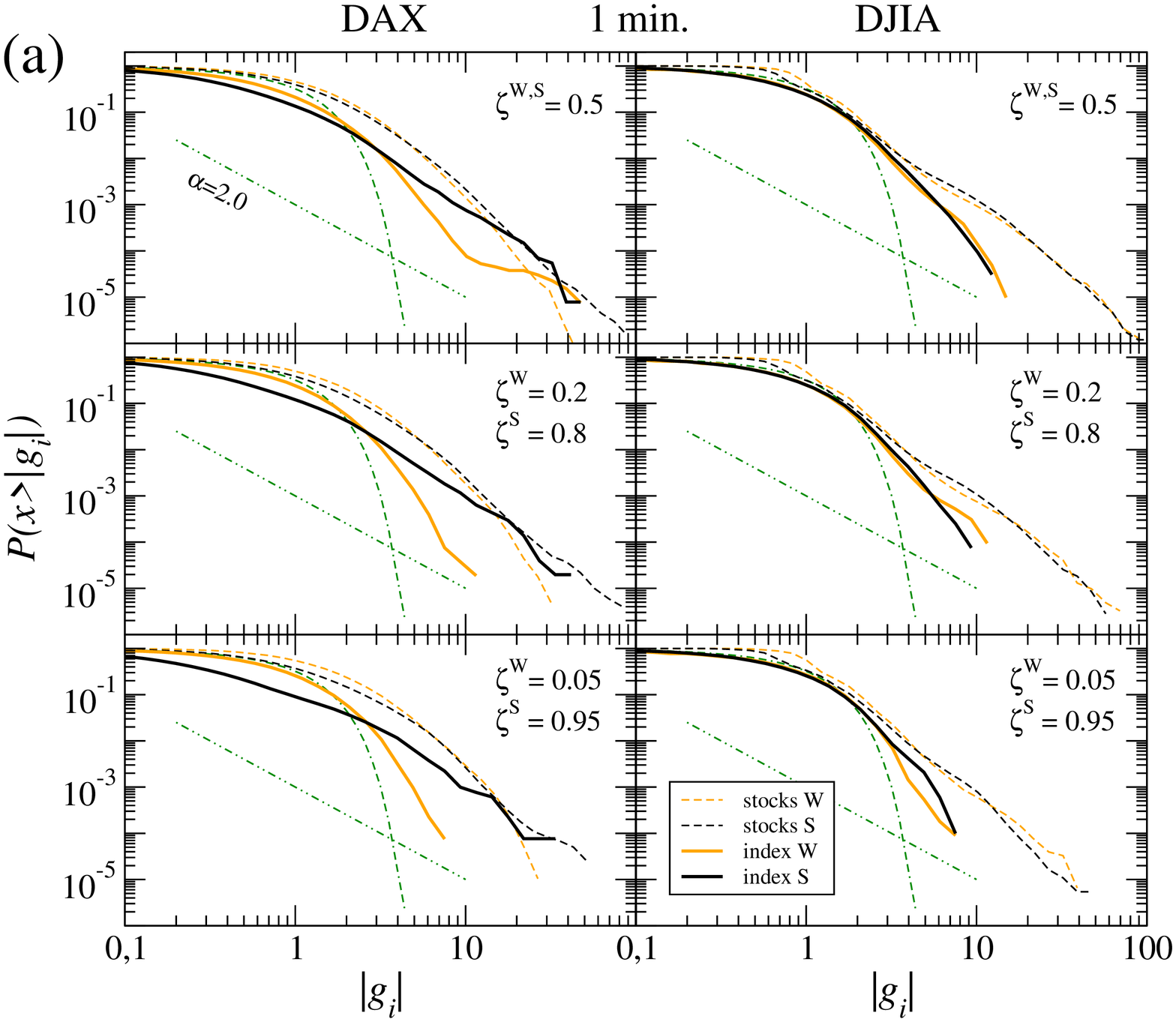}

\epsfxsize 11cm
\vspace{0.5cm}
\hspace{1.0cm}
\epsffile{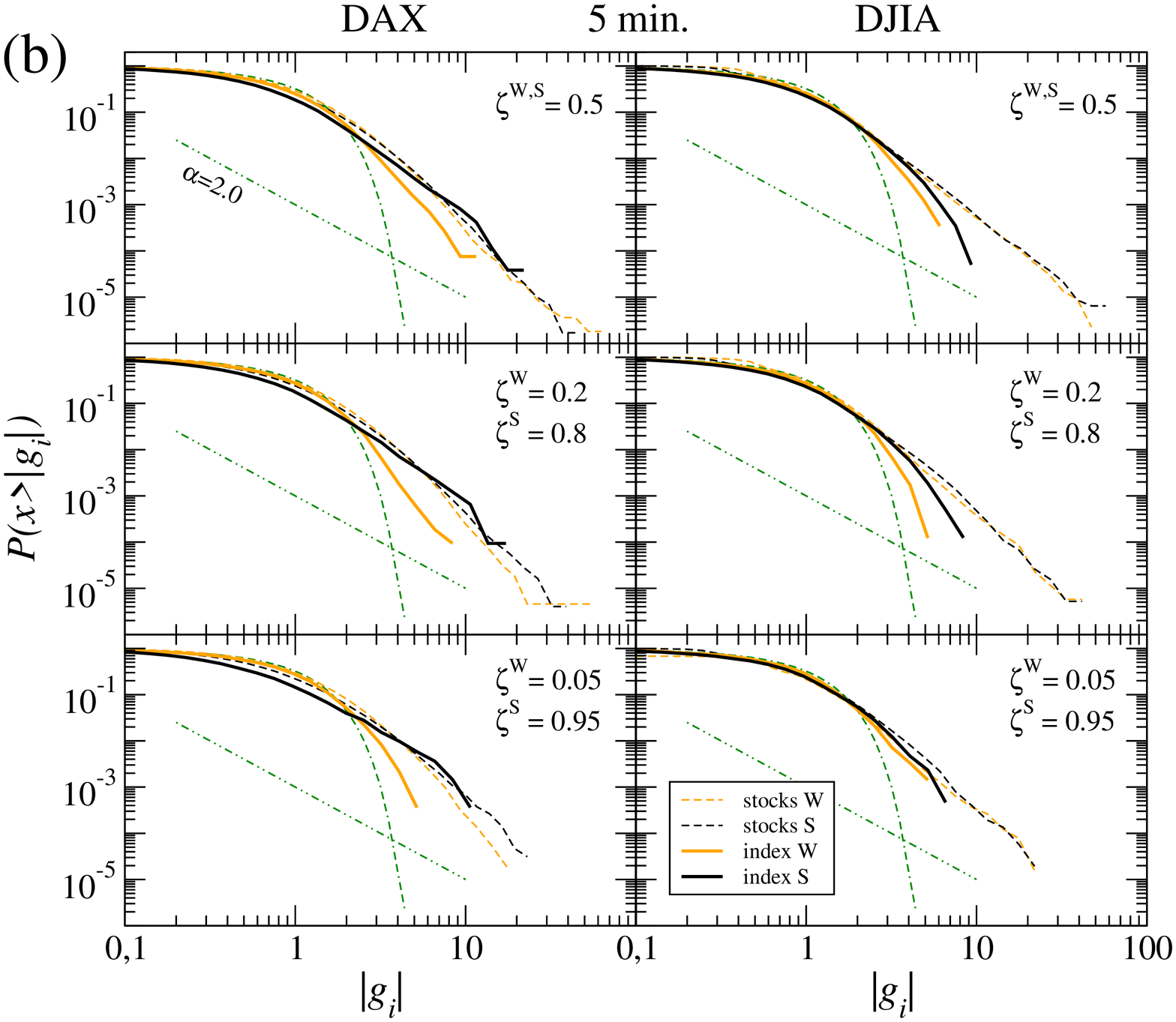}
\caption{Cumulative distributions of stock and index returns after  
the repeated normalization (see text) for the DAX (left) and the DJIA  
(right) markets and for four different time scales: 1 min (a), 5 min    
(b), 10 min (c) and 30 min (d). The distributions are denoted as in
Fig.~3. In each panel, the cumulative normal distribution is denoted by
a dash-dotted line while the L\'evy stable regime ($\alpha=2.0$) is   
denoted by a dash-double-dotted slanted line. In (c) and (d) no extreme 
distributions ($\zeta^W=0.05$ and $\zeta^S=0.95$) can be presented due
to too poor statistics.}
\end{figure}

\setcounter{figure}{3}

\begin{figure}
\epsfxsize 11cm
\hspace{1.0cm}
\epsffile{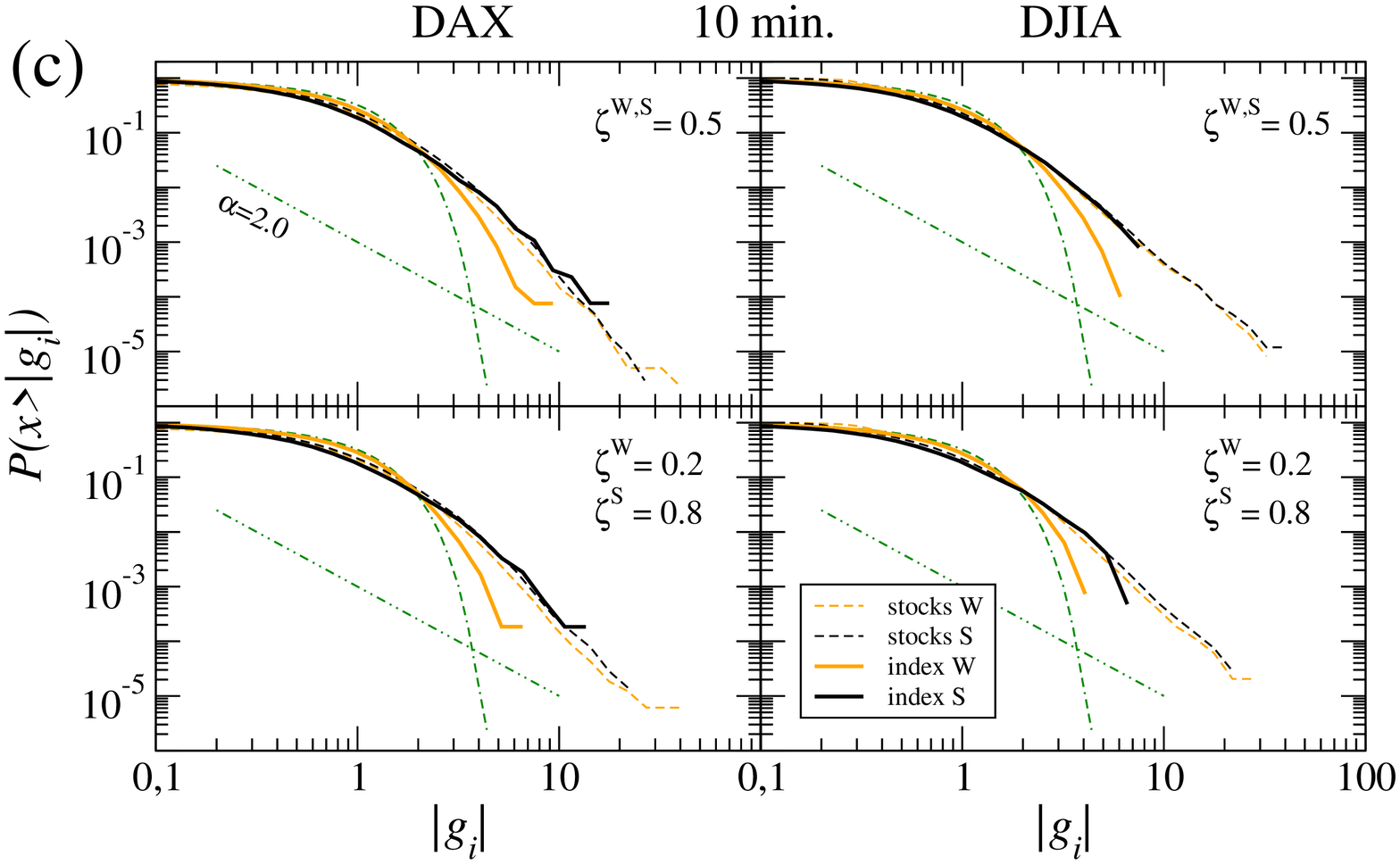}

\epsfxsize 11cm
\hspace{1.0cm}
\epsffile{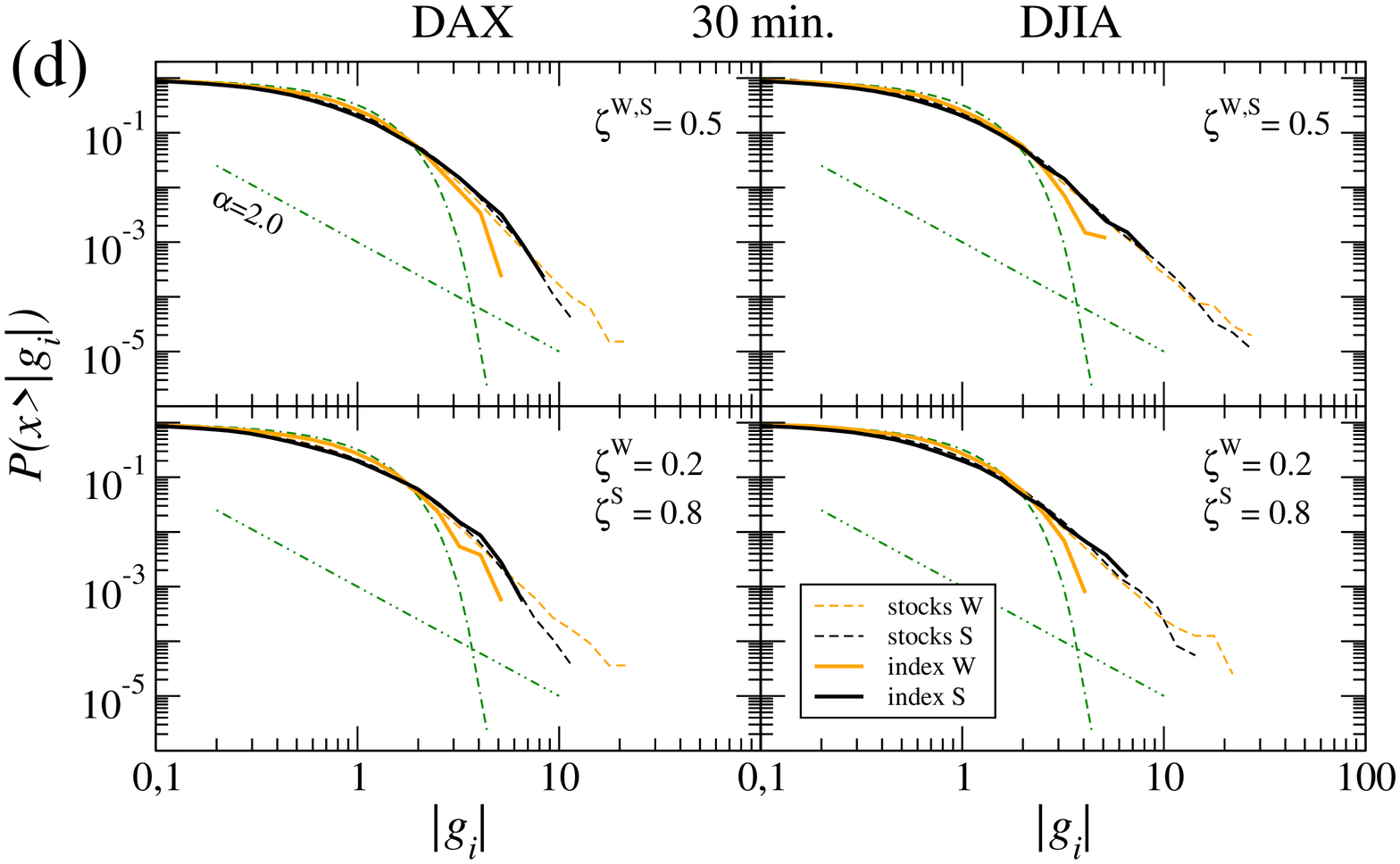}
\caption{Continued.}
\label{fig:fig4}
\end{figure}

The main conclusions which can be drawn from Figure 4 are as follows:

(i) The cumulative distributions of the stock returns (dashed lines) for
$W$ and $S$ windows (gray and black, respectively) reveal roughly similar
properties both in their central parts and the tails, regardless of the 
market, the time scale $\Delta t$ and the threshold separation $\Delta 
\zeta$. A perfect agreement of the distributions is evident for the DJI 
stocks, while some small differences can be observed for the German 
stocks, especially for short time scales.

(ii) On the contrary, the distributions of the index fluctuations (solid
lines) are heavily influenced by the strength of correlations between the
stock price movements. The distributions of the returns in $W$ windows
(grey solid line) tend to be more Gaussian than the ones associated with
$S$ windows (black solid line). This systematically applies to almost all
cases with an exception for $DJIA_I^W$ and $DJIA_I^S$ computed for $\Delta
t=1$ min, where the situation is less clear. Even in this case, however,
the two distributions seem to differ for the largest $\Delta \zeta$
(bottom right panel of Fig.~4(a)). For $\Delta t=5$ min and $\Delta
\zeta=0.9$ we do not observe so big difference between the distributions
in DJIA as we might expect, but this fact can well be attributed to poor
statistics (in this case, each of the two distributions was calculated
from 2100 returns only).

(iii) The larger $\Delta \zeta$ is, the larger is the difference between
$DAX_I^S$ and $DAX_I^W$; this rule is especially significant at small
$\Delta t$ and deminishes at larger $\Delta t$. For DJIA, the analogous
increase of separation is less pronounced (Fig.~4(a)-4(d)).

(iv) The above-mentioned increase of separation between $DAX_I^S$ and
$DAX_I^W$ with increasing $\Delta \zeta$ is associated with the
appearence of the distribution with scaling regions whose slope is in 
the L\'evy-stable regime. Yet another look at Fig.~2(a) allows one to see 
that even though the distributions of the DAX returns show fat tails,
they are by no means stable. A significantly different behaviour of the
distribution can be obtained after filtering out those returns that belong 
to the intervals of uncorrelated trading and whose distribution drops down
considerably faster (Fig.~4(a)). With the threshold as low as at
$\zeta^S=0.5$ (i.e. $\Lambda^S=5.76$), the $DAX_I^S$ distribution
presents the apparent scaling just at the edge of the L\'evy-stable
regime ($\alpha \simeq 2.0$, represented by the slant dash-dotted line
in each panel). By rising the threshold $\zeta^S \ge 0.8$ ($\Lambda^S \ge
7.90$) we observe the occurence and inflation of another scaling region
for smaller values of the returns with the scaling index deep inside the
stable range ($\alpha \simeq 1.3$). For $\Delta t=5$ min (Fig.~4(b)) we
also identify the scaling region but now with $\alpha$ continuously 
decreasing with increasing $\zeta^S$. This region is shorter than the 
one for $\Delta t=1$ min and falls into the stable regime at 
significantly higher threshold ($\zeta^S \simeq 0.8$). No scaling of 
$DAX_I^S$ can be observed for larger time scales (Fig.~4(c) and (d)).
Unlike DAX, DJIA  does not convincingly scale for any of the analyzed 
values of $\Delta t$ (compare ref.~\cite{gopi99}).

A striking difference between the properties of the DJIA and the DAX
returns distributions can be seen for the time scale of 1 min (Fig.~4(a)).
For DAX, the distributions for the $S$ and $W$ windows are disparate,
whereas for DJIA they have similar shape. One of the possible sources of
this discrepancy can be the already-mentioned more noisy evolution of the
DJI market. Indeed, our computation shows that if $\Delta t=1$ min, the
median of $P(\lambda_1(w_j))$ reaches only 4.22 for the DJI stocks; this
value does not stray much from the upper edge of the eigenvalue spectrum
of a Wishart matrix. Thus, even the relatively high values of
$\lambda_1(w_j)$ are not very distant from the random case, e.g. for the
middle right panel of Fig.~4(a), $\zeta^S=0.8$ corresponds to the
threshold $\Lambda^S=5.08$. This can well account for the fact documented
in the bottom right panel of Fig.~4(a) that the $DJIA_I^S$ and the
$DJIA_I^W$ distributions start to significantly differ only for
$\zeta^S=0.95$ ($\Lambda^S=6.40$). For comparison, the threshold values
$\Lambda^S$ for DAX stocks are as follows: 5.76 ($\zeta^S=0.5$), 7.90
($\zeta^S=0.8$) and 10.60 ($\zeta^S=0.95$). The strongest correlations and
large values of $\lambda_1(w_j)$ are more likely to occur at
market-specific periods of intraday
trading~\cite{liu99,drozdz01b,wang01,kwapien02}, especially in the German
market; such periods are usually associated with highly volatile behaviour
of the market and rapid collective movements of prices. An example can be
the sudden significant changes of DAX which frequently occur almost
precisely at 14:30~\cite{drozdz01b,kwapien02}. These changes are visible
predominantly on short time scales; on longer time scales they are often
averaged out and vanish. This can be considered as one of the possible
sources of larger collectivity of DAX on short time scales. Another source
is the strong influence of the NASDAQ and NYSE evolution on the Frankfurt
stock market as a whole. Daily pattern of $\lambda_1(w_j)$ fluctuations
reveals significant increase of the largest eigenvalue's magnitude after
15:30 (i.e. 9:30 in New York) when trading starts on the NYSE and NASDAQ
markets; also the 14:30 peak is caused by certain external factors. From
this point of view the American market can be considered as being
relatively independent and, thus, less correlated. However, as the big
difference between $DAX_I^W$ and $DAX_I^S$ (and between the distributions
of the DAX and the DJIA returns) cannot be attributed merely to the
strongly correlated dynamics of the German market, an explanation for the
observed shape of the distributions of the DAX returns requires more
extensive investigation.

\begin{figure}
\epsfxsize 11cm
\hspace{1.0cm}
\epsffile{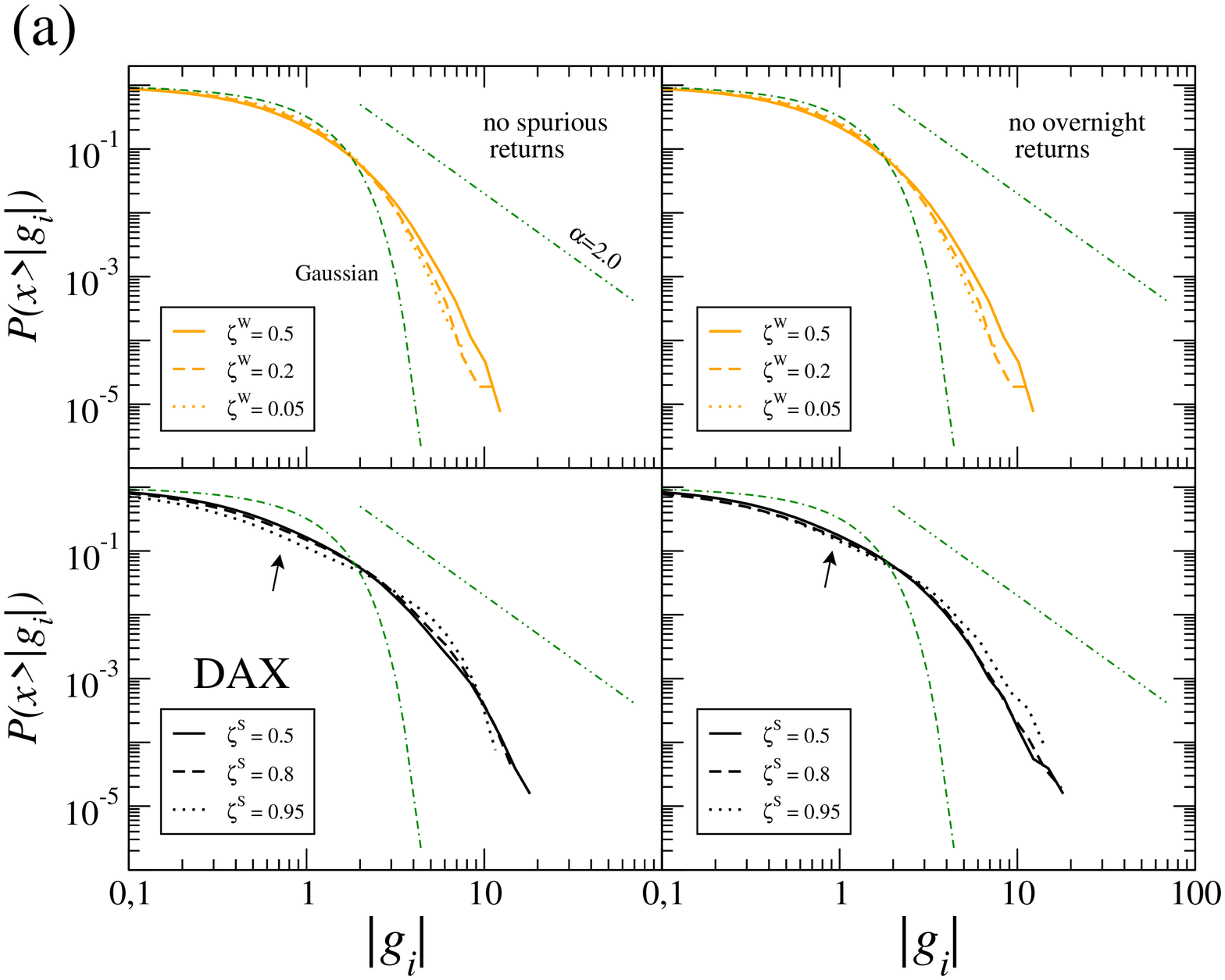}

\vspace{0.5cm}
\epsfxsize 11cm
\hspace{1.0cm}
\epsffile{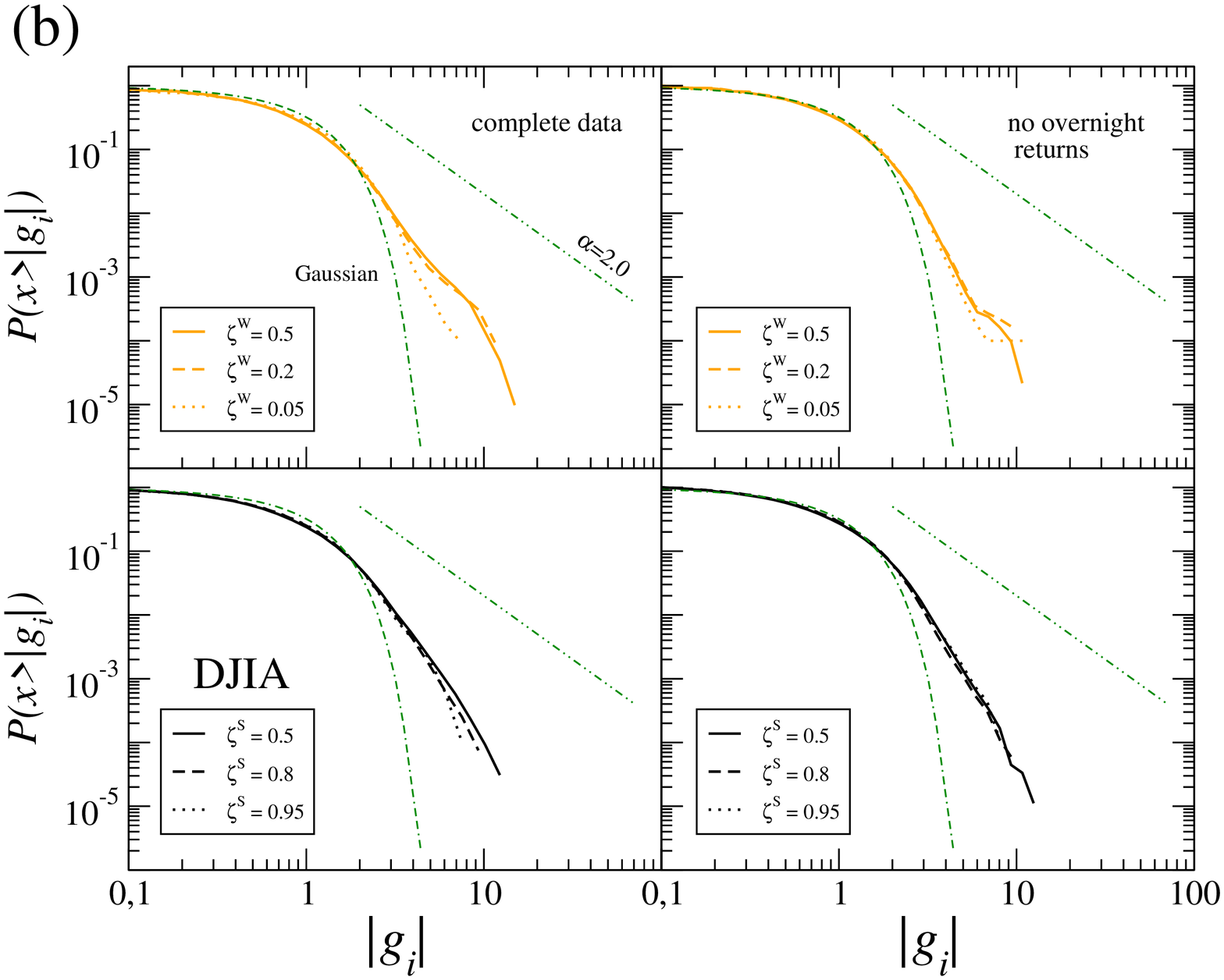}
\caption{Cumulative distributions of 1 min returns for DAX (a) and DJIA
(b). Three different realizations of the $W$ case (upper panels) and the 
$S$ case (lower panels) are presented. For DAX, panels on the left side of 
(a) present distributions calculated from data without spurious 
overnight returns, while panels on the right side present the same 
distributions after removing all the overnight returns. Left-hand side of 
(b) corresponds to complete DJIA data (i.e. the same as in Fig.~4(a)) 
and the right-hand side shows the same distributions but now comprising 
no overnight returns. The cumulative Gaussian and L\'evy stable 
distributions are also shown (dash-dotted lines). The small arrows in (a) 
point to the L\'evy scaling regions.}
\label{fig:fig5}
\end{figure}

A potential source of the fat-tailed distributions of the DAX returns may
be the calculation procedure of DAX after the market opening: an opening
value may be assigned to the index only after not less than 50\% of the
DAX component stocks representing at least 70\% of the total DAX market
capitalization have already been traded on a given day. This usually
happens a few minutes after the actual market opening and leads to a
significant relative amplifiction of the overnight DAX return if the
corresponding time scale is shorter than a few minutes (as it is in our
case of $\Delta t=1$ min). Obviously, for any longer time scale this
effect should not influence the magnitude of the corresponding returns; in
fact, Figs.~2(a), 4(c) and 4(d) show that the $S$ distributions for DJIA
and DAX look similar both for $\Delta t=10$ and for $\Delta t=30$ min. In
order to quantify the influence of the overnight DAX returns on the
corresponding distributions, we take our time series of 1 minute returns
of DAX and select all the returns which correspond to overnight changes of
the index value (i.e. the difference between a previous day's closing
value and the next day's opening value). Next we zero the spurious returns
corresponding to the situation in which the opening value was assigned to
DAX later than at 8:31:00 (as they should not be considered as valid 1 min
returns), and leave the remaining overnight returns unchanged (roughly
about half the total number of overnight returns). Then we calculate the
distributions according to the same procedure as for the complete time
series in Fig.~4(a). For a comparison, we also calculate the distributions
after zeroing all the overnight returns both the spurious and the valid
ones.

Figure 5(a) exhibits the cumulative distributions of the DAX returns taken
from the $W$ windows (top panels) and the $S$ windows (bottom panels)
separately; in each panel distributions for three different values of
$\zeta^{W}$ or $\zeta^S$ are shown. Both the distributions calculated from
the data without the spurious returns (on the left of Figure) and the
distributions calculated from the data without all the overnight returns
(on the right) have got significantly thinner tails than their
counterparts for complete data in Fig.~4(a); this is true both for the
collective and for the uncorrelated trading intervals. Moreover, the outer
scaling region with $\alpha \simeq 2.0$ which was clearly visible in
Fig.~4(a), here disappeared, which permits us to identify its origin as
being merely due to the calculation procedure of the DAX opening value.
The extremely fat tails for DAX observed in Fig.~4(a) may therefore be
considered as an external effect unrelated to the inner properties of the
market dynamics. These fat tails and scaling are more significant for the
$S$ windows than for the $W$ ones, but this is purely accidental: the
opening 30 minutes covered by the first window on each day is usually
associated with collective dynamics of stocks and thus it is classified as
an $S$ window.

\begin{figure} 
\epsfxsize 11cm 
\hspace{1.0cm} 
\epsffile{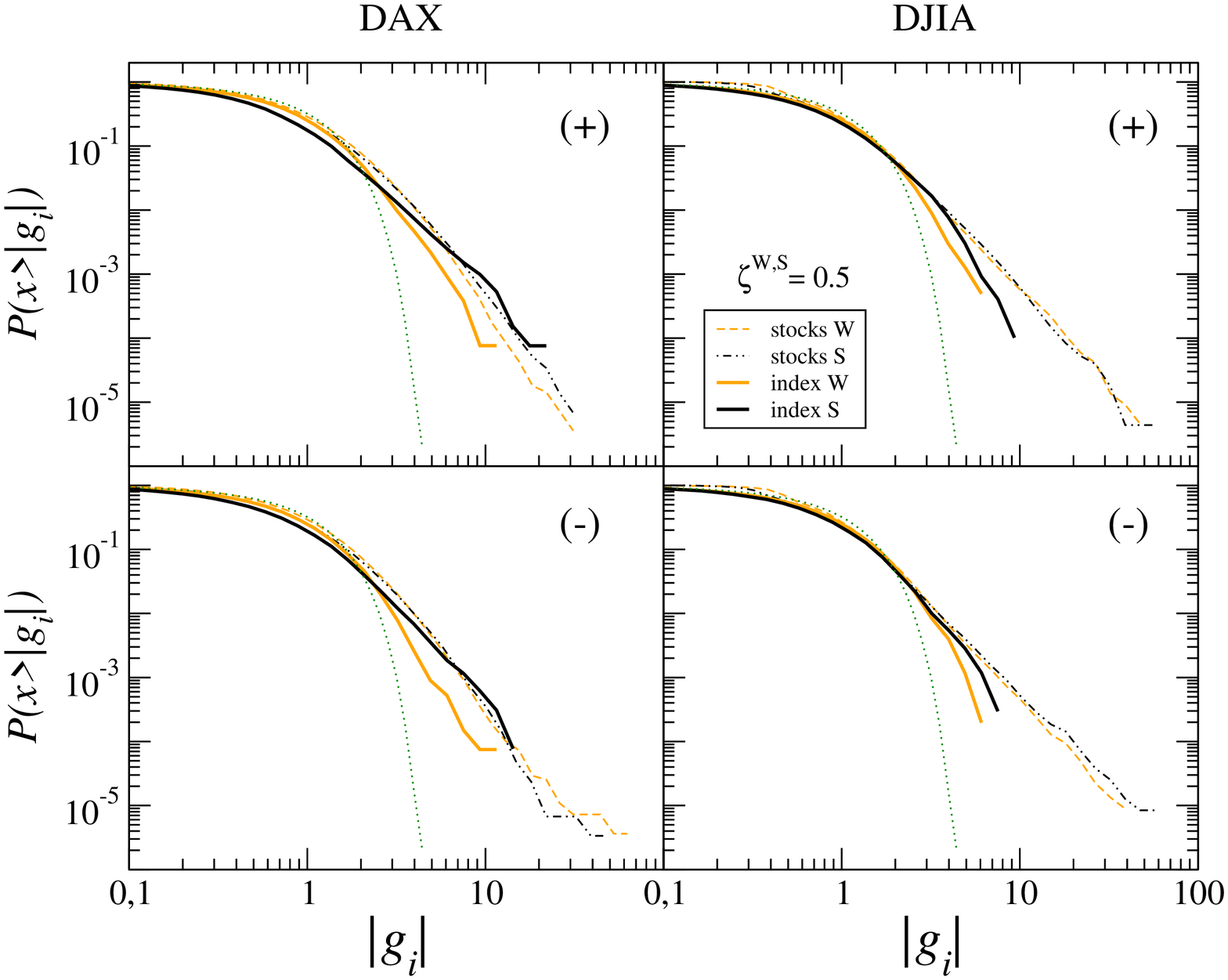}
\caption{Cumulative distributions of 5 min stock and index returns
(complete data and repeated normalization) for the DAX (left) and the DJIA
(right) markets. The upper (lower) panels correspond to positive
(negative) fluctuations. Only the case of $\zeta^{W,S}=0.5$ is shown. The
cumulative normal distribution is denoted by a dotted line in each panel.}
\label{fig:fig6} 
\end{figure}

In contrast to the outer scaling region with $\alpha=2.0$, the inner
scaling region characterized by $\alpha=1.3$ and which is best visible for
high values of $\zeta^S$ for $0.5<|g_i|<3$ (arrows in Figure), survives.
Removing the spurious returns does not affect it at all and removing also
the rest of overnight returns only slightly shorten it and change its
slope to 1.4. A similar region of scaling in the L\'evy-stable regime has
already been identified previously for the S\&P 500 index~\cite{gopi99};
in that case $\alpha=1.6$. Interestingly, a trace of a related feature
(but without clear scaling) can also be seen for the DAX-stocks returns in
Fig.~4(a) (bottom left panel, black dashed line). As this stable scaling
region occurs only in ``collective'' windows with large $\lambda_1(w_j)$
it seems that, for DAX, its origin cannot be attributed to the
quantization of the returns as authors of ref.~\cite{gopi99} hypothesize
for S\&P500. It is also interesting to note that after removing the
spurious returns, the distributions for the $S$-windows closely resemble
the distributions for the truncated L\'evy processes, where the central
part is a L\'evy distribution while, owing to the exponential cut-off of
tails, the second moment is finite (see ref.~\cite{mant95}). In fact, if
we calculate pdf of the price increments (instead of the log-returns) for
the $S$-windows, we obtain good fits by using L\'evy distributions with
$\alpha=1.3 \div 1.4$; the tails of such pdf's decay approx.~exponentially
with the best agreement being observed for $\zeta^S=0.8$. After removing
all the overnight returns, the tails do not present explicit exponential
decay but second moment is obviously still finite (bottom right panel of
Fig.~5(a)).

Figure 5(b) displays the results of an analogous analysis for DJIA. Due to
the fact that the opening values of DJIA may be calculated without any
restriction, there is no spurious returns and thus in Figure we show the
results for the complete data instead: the distributions on the left side
are exactly the same as those on the right side of Fig.~4(a). Even if we
remove all the overnight returns we do not observe any qualitative
modification of the distribution's shape (right side of Fig.~5(b)). Only
small quantitative changes can be seen here; without the opening returns,
the $DJIA_I^W$ distributions are characterised by slightly thinner tails
and start to differ from their $S$ counterparts already for $\zeta^S=0.8$
instead of $\zeta^S=0.95$ as it was for the complete data. This is not
shown explicitely but can be inferred from a careful inspection of the
right-side panels of Fig.~5(b). For DAX (Fig.~5(a)), the difference
between the distributions corresponding to $W$ and $S$ windows decreases
both after removing the spurious and the valid overnight returns, but it
is still stronger for DAX than for DJIA and increases with increasing
$\Delta \zeta$.

For longer time scales ($\Delta t \ge 5$ min), removing the spurious
overnight returns does not influence the results presented in
Fig.~4(b)-(d) for any of the two markets whereas removing also the valid
overnight returns influence the distributions. For $\Delta t \ge 10$ min,
the overnight returns share the properties with intraday returns and
removing them has no qualitative effect. Thus, we may conclude that the
large difference in the distributions of the returns between DAX and DJIA
observed in Fig.~4(a) is predominantly due to the following two factors:
(a) the properties of the overnight returns and especially the non-trivial
calculation of such returns in DAX, and (b) the different strength of
couplings in each of the two markets; the DJI stocks present more noisy
evolution than the German stocks, which leads to similarity of the
correlation properties of the $W$ and $S$ intervals and, in turn,
similarity of the corresponding distributions in DJIA.

Finally, we shall compare the properties of the distributions of the
positive and negative returns~\cite{gopi99,lillo00}. Such distributions
for $\Delta t=5$ min are displayed in Figure 6 both for the German (left)
and the American (right) markets. The positive fluctuations corresponding
to index drawups (upper panels) and the negative ones associated with
index drawdowns (lower panels) do not differ from each other
qualitatively, resembling the distributions for the absolute returns
presented for this time scale in Fig.~4(b). This is in agreement with the
findings of ref.~\cite{plerou99b,gopi99} that the properties of
distributions of the stock and index returns are symmetric with respect to
zero.

\section{Conclusions}

To summarize, the results of our analysis show that the time intervals
characterized by strongly collective behaviour of stocks are associated
with the distributions of the index returns, whose properties differ from
the ones for the intervals dominated by noise. Strongly correlated market
can be related to the phenomenon of fat tails of the returns distribution,
while faster convergence of such a distribution to normal can be
attributed to a decorrelated trading. This might be considered as an
empirical argument supporting the hypothesis stating that the important
factor responsible for the fat tails of the distributions of the index
returns is the inter-stock correlations~\cite{gopi99}. Such an effect is
observed in both the German and the American market for time scales of at
least 5 min and in the German market for even 1 min time scale. This does
not exclude, however, possible influence of other factors which can either
amplify the effect of inter-stock correlations or be a distinct source of
the non-Gaussian tails (see~\cite{lillo00}). For the DAX market, which is
in principle more collective than the Dow Jones one on short time
scales~\cite{drozdz00,drozdz02a}, the strong inter-stock couplings which
occur both repeatedly in specific periods of a trading day and uniquely at
random moments, lead to the occurence of the L\'evy-stable region in the
distributions of the index returns. This region, however, comprises
returns of moderate size only and its existence does not affect the
distributions' tails. The results of our study indicate that removing
spurious overnight returns in the $S$-windows leads to the distributions
which, on short time scales, closely resemble those for the truncated
L\'evy processes. The existence of and switching between different
fluctuation regimes in index evolution during periods of correlated and
decorrelated trading resembles the phenomenon of two-phase behaviour of
the demand for stocks where the equilibrium and the out-of-equilibrium
phase interweave~\cite{plerou03}. We do not observe, however, any sudden
change of properties of the fluctuations for any of the values of the
control parameter $\lambda_1$, but rather a continuous transition from one
type of behaviour to another type, which is best visible for 1 min and 5
min returns of DAX.

\end{document}